\documentclass[aps,prl,reprint]{revtex4-1}
\usepackage{graphicx}
\usepackage{mathrsfs}
\usepackage{subfigure}

\begin{document}

\title{Hamiltonian analogs of combustion engines: a systematic exception to adiabatic decoupling}

\author{Lukas Gilz}
\author{Eike Thesing}
\author{James R. Anglin}

\affiliation{\mbox{State Research Center OPTIMAS and Fachbereich Physik,} \mbox{Technische Univerit\"at Kaiserslautern,} \mbox{D-67663 Kaiserslautern, Germany}}

\date{\today}

\begin{abstract}
Workhorse theories throughout all of physics derive effective Hamiltonians to describe slow time evolution, even though low-frequency modes are actually coupled to high-frequency modes. Such effective Hamiltonians are accurate because of \textit{adiabatic decoupling}: the high-frequency modes `dress' the low-frequency modes, and renormalize their Hamiltonian, but they do not steadily inject energy into the low-frequency sector. Here, however, we identify a broad class of dynamical systems in which adiabatic decoupling fails to hold, and steady energy transfer across a large gap in natural frequency (`steady downconversion') instead becomes possible, through nonlinear resonances of a certain form. Instead of adiabatic decoupling, the special features of multiple time scale dynamics lead in these cases to efficiency constraints that somewhat resemble thermodynamics.
\end{abstract}

\maketitle

\section{I. Introduction}

\the\textwidth

\subsection{A. Adiabatic decoupling}
One of the most basic and broadly applicable principles in all of physics is \textit{adiabatic decoupling}. It is already familiar from school physics: two weakly coupled degrees of freedom will only exchange substantial amounts of energy if their resonant frequencies match. In more sophisticated language, adiabatic decoupling means that the effects of fast degrees of freedom on slow degrees of freedom can be described by renormalizing the effective Hamiltonian which governs the slow time evolution. Applications of this principle range from the Born-Oppenheimer approximation in molecular physics \cite{Born_Beweis_1928,Born_Zur_1927} to effective Langrangians and renormalization in quantum field theory \cite{Weinberg,Shankar_Renormalization_1994,Berges_Non_2002}. Whether the fast degrees of freedom in question are gluons in nuclear matter, atomic excitations, or vibrations in a rigid body, it is thanks to adiabatic decoupling that we can `adiabatically eliminate' or `integrate out' high-frequency modes, to leave an effective Hamiltonian which describes only slow variables, and yet describes them accurately, even over long times.

While the power of adiabatic elimination is often celebrated, its most basic implication is less frequently noted: since the slow evolution is Hamiltonian, it conserves energy. Fast degrees of freedom cannot steadily inject energy into the low-frequency world. Or can they?

\subsection{B. Steady downconversion}
Consider the case of combustion engines. These are such large and complex dynamical systems that they must in practice be described thermodynamically, rather than in microscopic terms. In principle, however, a real engine could still operate within a sealed container, at least for some time, as long as the container were large enough to include air, exhaust, and so on. The operation of a combustion engine is thus in principle a case of closed-system Hamiltonian time evolution. What a combustion engine nonetheless does, however, is to extract energy steadily from chemical degrees of freedom, whose dynamical time scales are on the order of femtoseconds (frequencies $\sim 10^{14}$ Hz), and then ultimately use this energy to turn wheels at a few thousand revolutions per minute ($\sim 10^{2}$ Hz). If we introduce the term \textit{steady downconversion} to mean this kind of steady transfer of energy across a dynamical frequency gap, precisely as is normally forbidden by adiabatic decoupling, then steady downconversion across a frequency ratio of $10^{12}$ is in fact achievable, albeit in dynamical systems too large and complex for microscopic description.

\subsection{C. Microscopic models}
The question of exactly how it is that macroscopic engines manage to evade adiabatic decoupling, and achieve steady downconversion, leads quickly into the general question of how thermodynamics `emerges' from mechanics in the limit of large system size. Much has been learned about that emergence, however, by starting from the microscopic side of the limit. The main goal of such work has hitherto been to learn how the ensembles of statistical mechanics can be justified by effective ergodization within Hamiltonian mechanics. As a fine example of this kind of study, we are inspired by the classic work of Chirikov, in showing how a not-too-small perturbation of an integrable system can break the invariant tori of Kolmogorov-Arnold-Moser theory \cite{Arnold_MMoCP_1989} and lead to effectively ergodic behavior \cite{Chirikow_1979}. 

In particular, Chirikov presented instructive models of nonlinear oscillation, and showed how certain resonant perturbations can induce Arnold diffusion \cite{Chirikow_1979}, even in systems with few degrees of freedom. In the present paper we take a similar approach, to the point of studying some quite similar Hamiltonians. The significant differences that we will introduce, however, will reflect a quite orthogonal perspective: our concern is not information, but energy. Instead of aiming to understand the effective loss of information that is represented in statistical mechanics by ensembles, we will focus on steady downconversion as a mechanical analog to the conversion of energy-dense fuel into work, as in a combustion engine. Our basic question is therefore: In a phase space of only a few dimensions, what kind of Hamiltonian system can exhibit steady downconversion? 

\subsection{D. Hamiltonian `daemons'}
Because our central concern is thus secular energy transfer between slow and fast degrees of freedom, it is an important restriction that we will further insist on a \textit{time-independent} Hamiltonian, so that the system conserves its total energy and no important energetic behavior is `put in by hand'. This restriction to time-independent Hamiltonians means that although we are only seeking a sort of minimal toy model for a combustion engine, yet in an important way our quest is more ambitious than most theoretical models for microscopic engines: we do not allow the kind of macroscopic elements that are represented by external control parameters (which would make the Hamiltonian time-dependent) or by heat reservoirs (which would make the evolution non-Hamiltonian). 

The kind of system we seek can thus be thought of as a sort of mechanical Maxwell's Demon --- except that instead of being directed against the Second Law of Thermodynamics, the principle it contrives to violate is that of adiabatic decoupling. It is also an important difference that our small Hamiltonian systems constitute our entire systems of study, and are not, like Maxwell's Demon, small additions to macroscopically large systems. Insofar as we are considering microscopic analogs of engines, we are considering \textit{entirely} microscopic analogs, not just microscopic working parts which extract energy from macroscopic heat baths. In our discussion at the end of this paper we will briefly compare this kind of isolated small system with systems coupled to baths.

To suggest the kinship of our systems with the Demon, while also acknowledging the important differences, we will refer more briefly to ``small dynamical systems with time-independent Hamiltonians which exhibit steady energy transfer from fast to slow motion'' as \textit{Hamiltonian daemons}. The first question of our paper is then: Do daemons exist, mathematically? The answer can be seen from Fig.~1. 
\begin{figure}[htbp]\label{fig1}
	\centering
{\includegraphics[width=.45\textwidth]{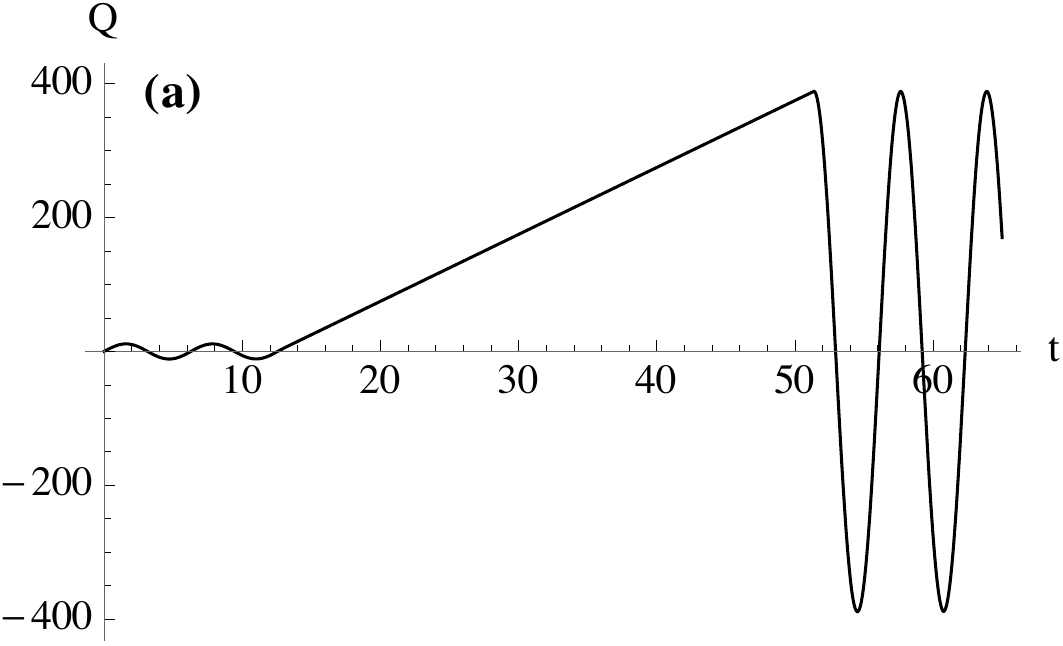}}
{\includegraphics[width=.45\textwidth]{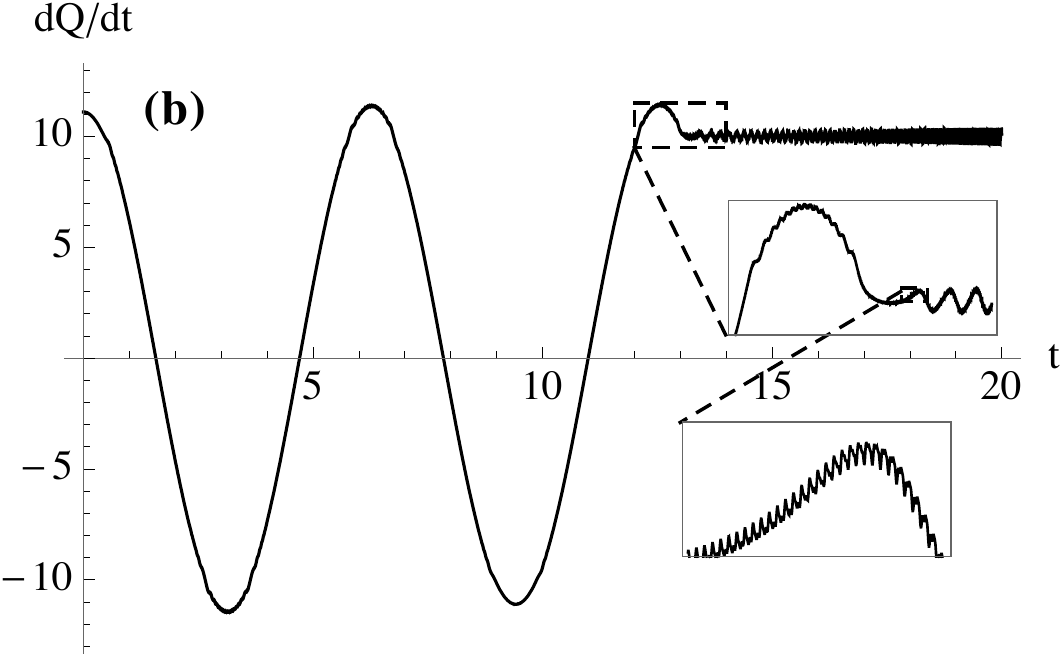}}
\caption{Evolution of a Hamiltonian daemon, under Hamiltonian $H_{1}$ of Eqn.~(1), with the parameter values stated in the text, and initial conditions $Q(0)=0,\dot{Q}(0)=11.1,q_{+}(0)=40/\sqrt{15},q_{-}(0)=p_{+}(0)=p_{-}(0)=0$. %
\textbf{(a)} Position $Q(t)$ of a mass. \textbf{(b)} Velocity $\dot{Q}(t)$, with insets showing detail. 
}
\end{figure}
\subsection{E. An example of a daemon}

What Fig.~1 shows is the position $Q$ (upper plot) and velocity $\dot{Q}$ (lower plot) of a certain point mass, which moves in time (horizontal axes) under the canonical equations of motion generated by the Hamiltonian (\ref{H1}), which we will present in Section II and then analyze for the bulk of this paper. Two other oscillators are also coupled in (\ref{H1}) to the $Q$ point mass, but are not represented in the Figure.

In Fig.~1 what we see is that $Q$ oscillates harmonically for a couple of periods, indicating that the point mass is moving in a quadratic potential. This point mass is also subject to other forces, however, and is not simply a harmonic oscillator, because after a few oscillations, it suddenly begins a `driving phase' of motion at nearly steady velocity, as if an engine had been turned on. After some time of this nearly steady motion, $Q$ abruptly resumes harmonic motion, and at a much larger amplitude than it had initially; something has done a lot of work on the mass during the driving phase of motion. In fact this work has come from the system's other two oscillators, which evolve on high frequencies. Small effects of these coupled high-frequencies are shown, in the insets of FIg.~1b), in the otherwise steadier speed of $Q$. The remarkable feature of Fig.~1, however, is a much more dramatic effect of the coupling of $Q$ to the high-frequency oscillators. During the driving phase of motion, when $\dot{Q}$ stays nearly constant, the high-frequency part of the system steadily transfers energy to $Q$.

Figure 1 is thus a proof by example that daemons are allowed by the laws of mechanics \cite{precision}. We will further show that daemons are a general class of dynamical systems featuring three distinct time scales, and that they obey certain non-trivial constraints, which are in some respects reminiscent of the laws of thermodynamics. We will approach these general points by first explaining the particular system of Fig.~1.

\section{II. A Hamiltonian Daemon}
\subsection{A. The Hamiltonian}
The Hamiltonian for the system of Fig.~1 is
\begin{eqnarray}\label{H1}
H_{1} &=& \frac{P^{2}}{2M}+\frac{M\nu^{2}}{2}Q^{2}+\frac{p_{+}^{2}+p_{-}^{2}}{2m}+\frac{m}{2}(\Omega_{+}^{2}q_{+}^{2}+\Omega_{-}^{2}q_{-}^{2})\nonumber\\
&&-Kq_{+}q_{-}\cos(k Q)\;,
\end{eqnarray}
with the particular values $M=500$, $\nu=1$, $m=1$, $\Omega_{+}=1500$, $\Omega_{-}=500$, $K=50\sqrt{3}$, $k=100$. Since $H_{1}$ does not depend explicitly on time, it is conserved. The work done on the mass depicted in Fig.~1 is drawn from the initial energy of the $\Omega_{+}$ oscillator, though the high frequency which actually matters dynamically in this system is really the difference $\Omega\equiv\Omega_{+}-\Omega_{-}=1000$. The quantitative parameters of this model were chosen for illustrative purposes; the system's behavior is not qualitatively sensitive to their precise values \cite{precision}.

Secular energy transfer in $H_{1}$ is non-trivial: if we let $K\to0$ we would simply have three harmonic oscillators, and even with the nonzero value of $K$ that we actually have, their three frequencies all differ by much more than the frequency scale $\sqrt{K/m}$ associated with the coupling between them. From this one would normally deduce \emph{adiabatic decoupling}: to zeroth order in $K$ the three co-ordinates should simply oscillate, and do so at greatly different frequencies, so that the coupling term would be not only small, but rapidly oscillating, and hence it should average to zero and have no significant effect of any kind. Beyond zeroth order, energy exchange would indeed occur, but it would be small in amplitude and high in frequency, with no net long-term (secular) effect. This kind of adiabatically decoupled behavior is indeed seen in Fig.~1 before and after the driving phase, but the driving phase defies the expectation of adiabatic decoupling.

The expectation of adiabatic decoupling in this case is based on the assumption that a weak coupling, such as we have, can only have substantial effects if it is also resonant. In linear dynamics, resonance means frequency matching, and since our three oscillators' frequencies are badly mismatched, linear dynamics would imply adiabatic decoupling. Weak but highly nonlinear couplings, however, may be resonant after all, at least in some regions of phase space, even across large frequency gaps. This is the kind of exception to adiabatic decoupling that our daemon exploits.

\subsection{B. Adiabatic decoupling vs. nonlinear resonance}
Numerical solution of the equations of motion for $H_{1}$ in fact confirms that adiabatic decoupling is the case throughout most of phase space. The coupling term in $H_{1}$ is a highly nonlinear function of $Q$, however, and this provides a nonlinear loophole in adiabatic decoupling. A special case occurs when $Q(t)$ happens to be rising or falling at close to the critical speed $v_{c}=\Omega/k=10$. In this circumstance, $q_{+}q_{-}\cos(kQ)$ is not just rapidly oscillatory, so as to average to zero, but includes a component which changes slowly. Whenever $\dot{Q}\sim \pm v_{c}$, therefore, the coupling may indeed have a more significant effect, at least for the short time during which the speed remains close to $v_{c}$\cite{explains}. 

An effect that is only significant for a short time is not really very significant, however; it would not count as an exception to adiabatic decoupling, which is about long-term behavior. But this is our key point, because as we will see below, when $\dot{Q}$ is near $v_{c}$, the significant effect of the coupling \emph{can be to keep $\dot{Q}$ near $v_{c}$}. So instead of a transient perturbation, the nonlinear resonance at $\dot{Q}=v_{c}$ can induce a transition to a new dynamical phase --- the driving phase. As Fig.~1 shows, $\dot{Q}$ can remain close to $v_{c}$ for a long time, until the slow mass has gained a large amount of potential energy from its steady climb up the harmonic potential $M\nu^{2}Q^{2}/2$. 

\subsection{C. Canonical transformation to $H_{2}$}
To see how the daemon works, we will apply canonical transformations and adiabatic approximations to transform $H_{1}$ into a series of effectively equivalent $H_{n}$. The first step is a canonical transformation to Hamilton-Jacobi variables $(\tau,U)$, $(\alpha,A)$ for the two fast oscillators:
\begin{eqnarray}\label{AA}
q_{\pm}&=&\frac{\sqrt{\pm 2(U-\Omega_{\mp}A)}}{\sqrt{m\Omega\Omega_{\pm}}}\cos(\Omega_{\pm}\tau+\alpha)\nonumber\\
p_{\pm}&=&-\frac{\sqrt{\pm 2m(U-\Omega_{\mp}A)\Omega_{\pm}}}{\sqrt{\Omega}}\sin(\Omega_{\pm}\tau+\alpha)\;.
\end{eqnarray}
If we then discard small terms that play no significant role in the Fig.~1 evolution because they are never resonant for $\dot{Q}$ near $+v_{c}$ (see our Appendix for details), we obtain the adiabatic effective Hamiltonian
\begin{eqnarray}\label{H2}
H_{2} &=& \frac{P^{2}}{2M}+\frac{M\nu^{2}}{2}Q^{2} + U\\
&& - \kappa\sqrt{(\Omega_{+}A-U)(U-\Omega_{-}A)}\cos[k(Q-v_{c}\tau)] \nonumber\\
\kappa &=&\frac{K}{2m\Omega\sqrt{\Omega_{+}\Omega_{-}}}=5\times 10^{-5}\hbox{ in Fig.~1}\;.\nonumber
\end{eqnarray}
Since the angle variable $\alpha$ does not appear in $H_{2}$, its conjugate momentum $A$ is a constant of the motion under the adiabatically approximate evolution generated by $H_{2}$ (hence an adiabatic invariant in the exact $H_{1}$ evolution). We may therefore consider $A$ as a constant, whose value is given by $A(0)$, in the Hamiltonian evolution of $Q$, $P$, $\tau$ and $U$. Inverting (\ref{AA}) yields $A(0)=8\times 10^{4}$ in the case of Fig.~1. 

\subsubsection{Comparison with a Chirikov resonance}
(Readers who are not interested in the relationship between our model and an isolated Chirikov resonance may skip directly to the next subsection.)

The coupling term in $H_{2}$ of (\ref{H2}) is now similar in mathematical form to the simplest case (an isolated resonance) of the resonances studied by Chirikov in \cite{Chirikow_1979}. One apparently significant difference, however, is that our $\tau$ is a canonical variable, which evolves under the time-independent Hamiltonian of the total system, whereas the corresponding variable (also called $\tau$) in Ref.~\cite{Chirikow_1979} is simply a multiple of the time, and appears as the argument of a sinusoidally time-dependent external perturbation. This difference is not really as sharply qualitative as it might seem, though, because as Chirikov points out, a fixed time dependence of the form $\tau\propto t$ can always be regarded as the evolution of a dynamical variable like our own $\tau$, if a term linear in $\tau$'s conjugate variable $U$ is added to the Hamiltonian. As long as $U$ appears nowhere else in the Hamiltonian, then, the equation of motion for $\tau$ will simply be $\dot{\tau}=1$. 

With this understanding (and after minor adaptations of notation) we can therefore express the Chirikov Hamiltonian of Eqn.~(3.15) in Ref.~\cite{Chirikow_1979} as
\begin{eqnarray}\label{Chirikov315}
H_{Ch}=K(I) + U+ G(I)\cos(k\theta-\Omega\tau)\;,
\end{eqnarray}
where $\theta,I$ and $\tau,U$ are canonically conjugate pairs. Thus the first difference between our model and Chirikov's is that the factor which multiplies our cosine coupling depends on $U$ instead of $I$. We will see in Section IV, below, that this dependence on $U$ is important in ensuring that the daemon's downconversion must eventually cease if $U$ is bounded from below (\textit{i.e.} our fuel source is finite). For the ongoing operation of the downconversion mechanism, however, this detail of coupling dependence on $U$ is not really significant, because our $U$ will only change slowly.

The most basic difference between our $H_{2}$ and the above $H_{Ch}$, in fact, is that in $H_{Ch}$ the slow variable $\theta$ which is coupled to the fast $\tau$ does not appear in $K(I)$, which depends only on the conjugate variable $I$. In our case, in contrast, the variable coupled to $\tau$ is $Q$, and the slow mass's own Hamiltonian includes not only a term that depends on its conjugate, $P$ (namely the kinetic energy $K(P)=P^{2}/(2M)$), but also a term that depends on $Q$ itself (the slow-sector potential energy $V_{S}(Q)=M\nu^{2}Q^{2}/2$). As we will explain just below, the precise form of this potential term in our model is not important for the downconversion mechanism; all that really matters is that we have a sufficiently slowly varying potential $V_{S}(Q)$. In this sense the Chirikov resonance is a degenerate case of our daemon model, in which $V_{S}(Q)$ happens to vanish. As we will see further on in this Section, the existence of a non-trivial potential term makes a qualitative change in our model's dynamics, in comparison with that of the isolated Chirikov resonance which our model otherwise resembles. The additional potential term introduces an additional time scale, and the phenomena that occur on this long additional time scale are the main subject of this paper: the long additional time scale corresponds precisely to the rate of steady downconversion. In the Chirikov limit $H_{Ch}$, this rate goes to zero. 

\subsection{D. Slowly varying potential}

The harmonic potential $M\nu^{2}Q^{2}/2$ changes only very slightly over the period of the rapid oscillations in $\dot{Q}(t)$ that occur during the driving phase. We can therefore understand the dynamics in this phase by replacing the harmonic potential with a linear potential whose slope $g$ is constant, and then after this simplified problem has been solved, replacing $g$ with the slowly changing slope $M\nu^{2}Q(t)$ of the actual harmonic potential. This yields
\begin{eqnarray}\label{H3}
	H_{3}&=&\frac{P^{2}}{2M} + g Q + U \\
&& - \kappa\sqrt{(\Omega_{+}A-U)(U-\Omega_{-}A)}\cos[k(Q-v_{c}\tau)]\;. \nonumber
\end{eqnarray}
In fact we could apply a similar procedure to replace any sufficiently slowly varying $V_{S}(Q)$, not necessarily quadratic, with an instantaneous linear potential whose slope changes slowly in comparison with the time scale that is relevant for the downconversion mechanism. We have confirmed with numerical solutions of such nonlinearly modified alternatives to $H_{1}$ that this procedure accurately approximates the exact evolution.

\subsection{E. An explicitly time-dependent invariant}

We now reach a crucial point in daemon dynamics. Linearizing the harmonic potential of $H_{2}$ has given the new Hamiltonian $H_{3}$ a first integral: 
\begin{equation}\label{Jdef}
J=U+(P+gt)v_{c}
\end{equation} 
is constant (even though it depends explicitly on $t$!) under the evolution generated by $H_{3}$. In the adiabatically decoupled phase, $J$ stays constant by having $\dot{U}\doteq 0$ and $\dot{P}\doteq -g$; in the driving phase it is constant because $\dot{U}\doteq -gv_{c}$ and $\dot{P}\doteq 0$. For the evolution shown in Fig.~1, $J\doteq 1.2\times 10^{8}$.

Using the first integral $J$ we can reduce $H_{3}$ to a single non-cyclic degree of freedom by a further canonical transformation from $(Q,P),(\tau,U)$ to $(q,p),(\tau,J)$, with $q = Q-v_{c}\tau$ and $p = P-M v_{c}$. With the explicit $t$-dependence of $J$, this is a time-dependent canonical transformation, and the transformed Hamiltonian which results \cite{Goldstein} is
\begin{eqnarray}\label{H4}
H_{4}&=& \frac{p^2}{2M} + gq - \kappa\sqrt{(\Omega_{+}A-\tilde{U})(\tilde{U}-\Omega_{-}A)}\cos(kq)\nonumber\\
\tilde{U}(p,t) &=& J-(p+Mv_{c}+gt)v_{c}\;.
\end{eqnarray}
Since $J$ is constant, we can now treat $H_{4}$ as a time-dependent Hamiltonian in the two-dimensional phase space $(q,p)$. Note in comparison that a pure Chirikov resonance, with $g\to0$ as in $H_{Ch}$ above, would have a time-independent Hamiltonian in place of $H_{4}$, as in the `resonance Hamiltonian' $H_{r}$ of Eqn.~(3.18) in Ref.~\cite{Chirikow_1979}.

\subsection{F. A `tilted washboard' model}

Finally, the $p$-dependence of $\tilde{U}$ can be important during the dynamical transition from decoupling to driving, but during bound motion the kinetic term in $H_{4}$ must necessarily be smaller than the depth of the local potential minimum, and so $p\lesssim\sqrt{M\kappa J}$. This means that the proportional contribution of $pv_{c}$ within $\tilde{U}$ is $\mathcal{O}(\sqrt{\kappa Mv_{c}^{2}/J})\sim 10^{-5}$. We can therefore understand both the driving and decoupled phases by neglecting $pv_{c}$ in $\tilde{U}$, obtaining for our final approximate adiabatic Hamiltonian simply that of a single particle in a time-dependent potential:
\begin{eqnarray}
H_{5} &=& \frac{p^{2}}{2M}+V(q,t)\nonumber\\
V(q,t)&=&gq - \kappa \sqrt{u(t)[\Omega A-u(t)]}\cos(k q)\nonumber\\
u(t)&=&J-\Omega_{-}A-Mv_{c}^{2}-gv_{c}t\equiv u_{0}(1-\gamma t)\;.
\end{eqnarray}
We emphasize that the time-dependent effective potential $V(q,t)$ which appears in the final approximate $H_{5}$ is not the same as the original slowly varying external potential $V_{S}(Q)=M\nu^{2}Q^{2}/2$ from the full and exact $H_{1}$. See our Appendix for numerical confirmation that $H_{5}$ accurately reproduces the exact evolution under $H_{1}$, as shown in Fig.~1, except for oscillations on the short $1/\Omega$ time scale.

Inserting the adiabatic time dependence $g = M\nu^{2}Q(t)\doteq 500 \times 10(t-12.5)$ reveals that the rate $\gamma$ in $u(t)$ as just defined is of order $10^{-2}$ throughout the evolution shown in Fig.~1. This is by far the lowest rate or frequency in the entire evolution. The effective potential $V(q,t)$ can thus be considered adiabatically, by understanding the motion implied by $V(q,t)$ for fixed $t$, and then allowing the parameters in that motion to change slowly with $t$. For any fixed $t$, $V$ has the form of a tilted sinusoidal `washboard'; see Fig.~2. The two dynamical phases of $H_{1}$ are then finally clear: for most $q,p$ orbits the slight undulations of the potential are insignificant (adiabatic decoupling), but as long as the washboard tilt is not too great, there exist local minima in $q$ of $V(q,t)$, about which bound motion is possible in the surrounding local potential well. 

When $p$ thus oscillates around zero, $P$ oscillates slightly around $Mv_{c}$, and hence $\dot{Q}$ oscillates slightly around the steady speed $v_{c}$, as seen in Fig.~1 for $13\lesssim t\lesssim 51$ (the driving phase). (These small oscillations of $q,p$ in the local well of $V$ are the intermediate-scale oscillations seen in the upper inset of Fig.~1b).) The crucial point is that by maintaining the mass's speed near $v_{c}$, the rest of the system steadily does work on the mass, by raising it steadily against the $gQ$ or $M\nu^{2}Q^{2}/2$ potential. This is how the daemon performs steady downconversion.

\begin{figure}[t]\label{fig3}
\includegraphics[width=.45\textwidth]{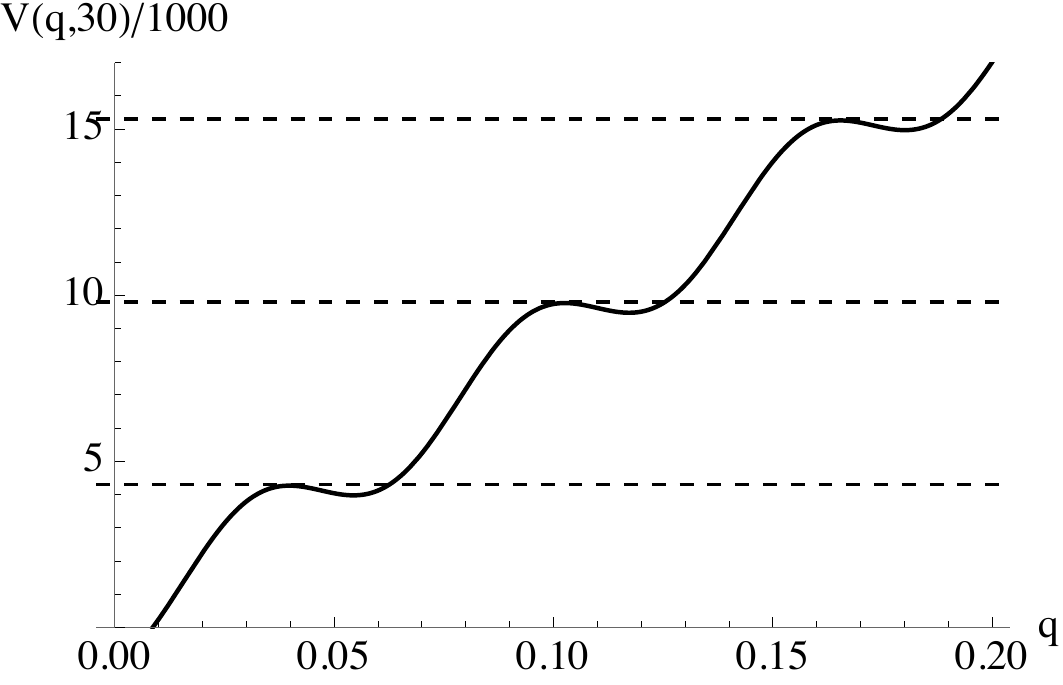}
\caption{Effective potential $V(q,t)$ from ${H}_{5}$, which represents $H_{1}$ in new canonical variables, with linearized external potential and adiabatic elimination of non-resonant terms. The time $t=30$ is chosen for illustration; the potential looks similar for other $t$ during the driving phase of Fig.~1. Dashed lines are horizontal guides to the eye, to confirm that shallow wells exist. Unbound orbits correspond to the adiabatically decoupled phase of $H_{1}$; bound orbits, to the steady downconversion of the daemon's driving phase.}
\end{figure}

\subsection{G. Self-sustaining resonance}
The simple picture of effective particle motion that we finally reached in $H_{5}(q,p,t)$ does let us understand the daemon's driving phase in familiar terms, but only through the rather abstract variables that emerged after a series of transformations and adiabatic approximations. The Hamiltonian daemon represented by $H_{1}$ is not a trivial system.

One somewhat more physical way to understand the real nature of the steady downconversion mechanism is to consider it as a nonlinear version of the ordinary kind of resonant energy transfer, with the special nonlinear feature being that the resonance only holds for $\dot{Q}$ close to $v_{c}$. As usual, therefore, energy transfer will be most efficient at exact resonance, which occurs when $\dot{Q}$ is exactly $v_{c}$. The mass might therefore rise stably at a speed just above $v_{c}$, so that if it were to decelerate, transfer efficiency would rise and the additional force would speed the mass back up. Conversely, if it were to accelerate, transfer efficiency would fall, and the mass would slow down. The nonlinear resonance can therefore be self-sustaining.

This effect is an example of what is generally known as \textit{nonlinear stabilization} \cite{Chirikow_1979}; but the effect of the stabilization is here somewhat different from usual. Ordinarily what is meant by nonlinear stabilization is that nonlinear effects \textit{prevent} the secular transfer of energy between degrees of freedom. In a Chirikov resonance as described by $H_{Ch}$ in (\ref{Chirikov315}) above, for example, a linear $K(I) = \omega I$ would mean that resonance ($k\omega = \Omega$) would obtain, or not obtain, throughout the whole phase space uniformly, depending only on the fixed parameters in the Hamiltonian (namely $k\omega$ and $\Omega$). If resonance does hold, therefore, the special case of linear $K(I)$ allows complete transfer of energy from $U$ to $I$. As well as being special, the case of linear $K(I)$ is only a degenerate case of downconversion, because the designation of $\tau$ as fast and $\theta$ as slow is in this case arbitrary: one could give the two oscillators the same frequency by the canonical rescaling $\theta = X\tilde{\theta}$, $I = \tilde{I}/X$ for $X=\omega/\Omega$. 

If $K(I)$ in $H_{Ch}$ is nonlinear, however, then the resonance condition $k K'(I)=\Omega$ will generally only hold for a particular value of $I$ --- just as in our case the quadratic kinetic energy $P^{2}/(2M)$ means that resonance occurs only at $P=Mv_{c}$. If $I$ is then increased in the Chirikov model, due to resonant transfer from $U$, resonance no longer holds, and transfer stops. The result in the Chirikov model is that $I$ can then only oscillate around the resonant value, with an amplitude limited by the strength of the cosine perturbation amplitude $G(I)$; the nonlinear stabilization prevents the secular transfer of energy from $U$ to $I$. In our case also the resonant transfer phase of evolution (the driving phase) sees only small oscillations of $P$ around the resonant value $Mv_{c}$, and the amplitude of these oscillations is limited by our interaction strength $\kappa$. Since our slow mass's energy also includes a potential term $V_{S}(Q)\to gQ$, however, the nearly steady maintenance of $P\doteq Mv_{c}$ means that the potential energy steadily increases, through resonant transfer from $U$. In our daemon model, therefore, the nonlinear stabilization of $P$ therefore means precisely that secular energy transfer is not prevented, but sustained. 

This is the basic mechanism of steady downconversion in Hamiltonian daemons like the one described by our $H_{1}$. The important consequence is that total energy transfer from $U$ to the slow mass is not limited by the strength of the coupling term, but by the total amount of available $U$ (and by other constraints due to the slow $\gamma$-time-scale time dependence of $H_{4}$, to be described below). The strength of the coupling term does limit the daemon's performance; but it limits it rather in the way that a car engine's performance is limited by its maximum torque output. A car whose engine is too weak cannot climb too steep a hill. If the torque suffices for the hill's steepness, however --- corresponding here to the tilt of $V(q,t)$ being gentle enough, in relation to the well depth, that local minima still exist --- then the car can keep driving uphill until it runs out of fuel.

\section{III. Generic Daemons}

The (mathematical) existence of Hamiltonian $H_{1}$ proves that Hamiltonian daemons exist. How representative is $H_{1}$ of daemons in general? In a word, broadly. It is representative of all daemons whose slow `work' degree of freedom is powered from high-frequency `fuel' that is not in itself chaotic, and which exploit isolated nonlinear resonances. For all such systems, the high-frequency sector can be represented without loss of generality with Hamilton-Jacobi co-ordinates $\tau,U$, plus as many additional co-ordinates $\vec{\alpha}$ as may be needed for all the high-frequency degrees of freedom. Since the Hamilton-Jacobi construction makes the $\vec{\alpha}$ time-independent under the evolution generated by $U$ alone, the $\vec{\alpha}$ can only evolve under the Hamiltonian terms which couple $U$ to the slow degree of freedom. These terms have to be small, because we can only identify the slow and fast degrees of freedom as such by overlooking their coupling on the grounds that it is weak. The $\vec{\alpha}$ variables will therefore evolve slowly under the full daemon $H$, and so considerations of resonance and adiabatic invariance like those applied to $H_{1}$ will also lead, for a large class of daemon systems, to explicitly time-dependent adiabatic invariants like $J$, and thus to time-dependent effective Hamiltonians of the forms of $H_{4}$ and $H_{5}$. The potential function $V$ may be different in its dependence on $u$ as well as on $q$, and it may have further slow time dependence through its dependence on $\vec{\alpha}$ variables. The basic scenario, of steady downconversion maintained by cyclic orbits in an effective potential, will remain the same. 

In particular it is interesting to note that daemons are generically dynamical systems with at least three important time scales. By their definition they have an important short time scale, analogous to $\Omega^{-1}$ in $H_{1}$. Also by the definition of steady downconversion, they must have a much longer time scale, analogous to $\gamma^{-1}$ in $H_{5}$, over which the high-frequency energy is transferred into slower motion. The steady slowness of this transfer is ultimately also due to the smallness of the interaction between the slow and fast subsystems; if it were not so slow, then the energy transfer would not really be analogous to the steady downconversion of an engine, but to the explosively non-steady downconversion of a bomb.

It is the intermediate third time scale which is surprising here: the period of effectively bound motion during the driving phase, as shown in the upper inset of Fig.~1b. The generic requirement for (adiabatically) periodic motion of this kind is clear from the nature of dynamical stabilization: Hamiltonian systems maintain a steady average speed by oscillating around it. If we label the oscillation frequency with $\omega$, then the weakness of coupling must imply $\omega\ll\Omega$, while the steadiness of downconversion implies $\omega\gg\gamma$. Daemons are thus examples of multiple-time-scale systems in which three distinct time scales are simultaneously important. Whether by coincidence or not, the essential feature of steady periodic motion in a daemon's driving phase is analogous to the cyclic operation of a macroscopic engine. We represent this fact provocatively in Fig.~3 by comparing $\Omega$, $\omega$, and $\gamma$ to the analogous rate scales of a motor.

Daemons are thus a large class of systems, but very many of them can be mapped onto systems similar to the one presented here. The exploration of possible exceptions and generalizations must be left for future work. Here we will turn instead to examining an important general feature of daemons which can also be understood from the behavior of $H_{1}$. How and why do daemon driving phases end?
\begin{figure}[t]\label{fig2}
	\centering
{\includegraphics[width=.3\textwidth]{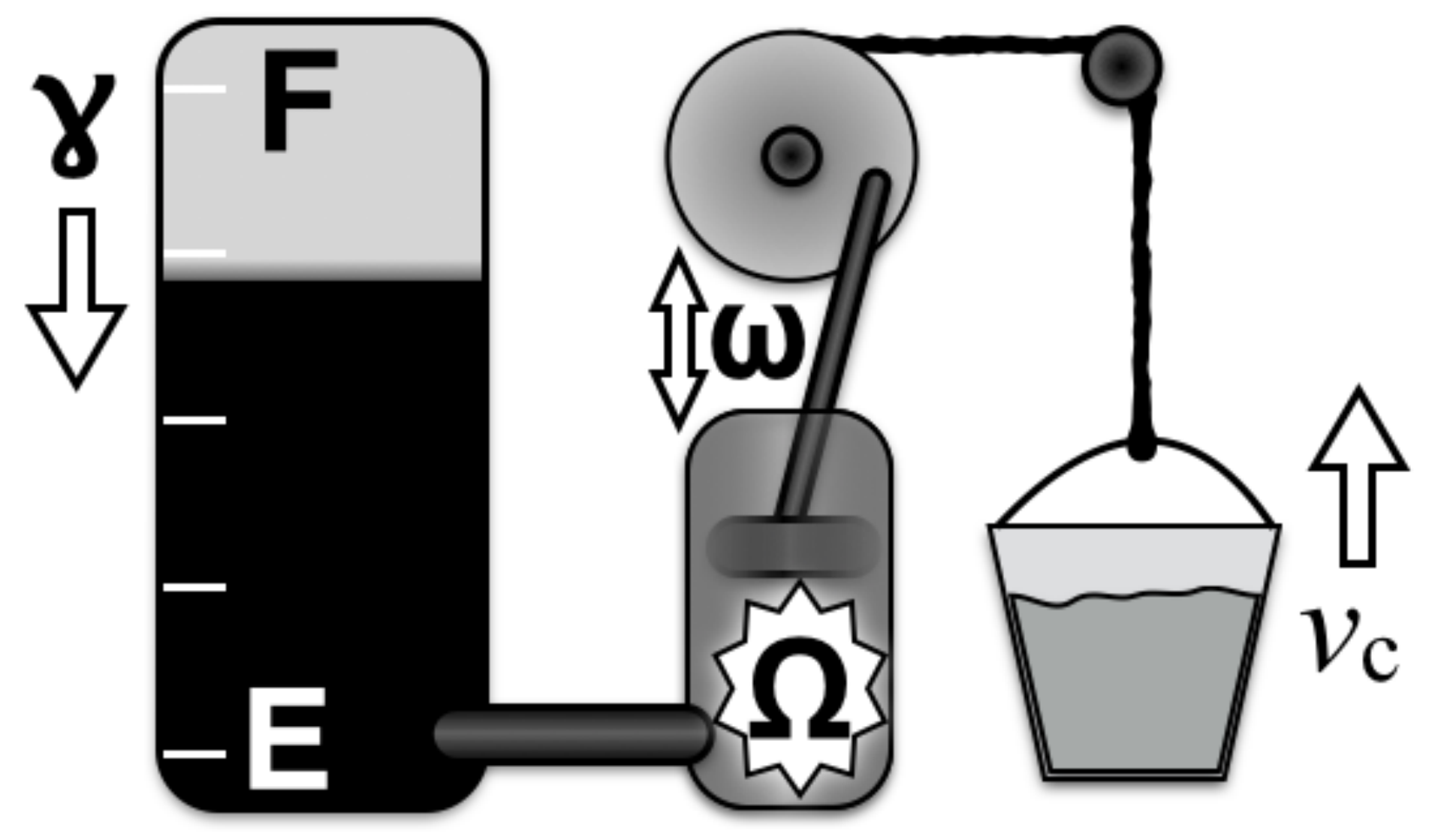}}
\caption{Schematic: Daemons as systems with three time scales, analogous to the dynamical time scales of a combustion engine. Energy is stored densely in degrees of freedom with a high dynamical frequency $\Omega$. This energy is consumed at a slow rate $\gamma$ to do work at a steady speed $v_{c}$. The fact that $\gamma\ll\Omega$ distinguishes engine-like dynamics (steady downconversion) from bomb-like (rapid energy release). The non-trivial mechanism which achieves steady downconversion involves cycles on the intermediate frequency scale $\omega$, with $\gamma\ll\omega\ll\Omega$.}
\end{figure}

\section{I.: Efficiency Limits}
\subsection{A. Energy conservation}
The explicit slow $\gamma t$ time dependence of $H_{5}$ is \emph{not} any kind of heuristic modification of the system dynamics, but is due entirely to the time-dependent transformation which replaced $U$ with $J$. That step, which produced $H_{4}$ from the time-independent $H_{3}$, was an exact canonical transformation. Since as a result $H_{5}$ is explicitly time-dependent, it is not conserved; but the original energy $H_{1}$ still is conserved. Indeed $H_{5}$ is time-dependent precisely \emph{because} $H_{1}$ is conserved, since the explicit time dependence of the adiabatic invariant $J=U+(P+gt)v_{c}$ reflects the fact that if $P\doteq Mv_{c}$ remains (nearly) constant despite the external force $-g$, then $U$ must decrease steadily to supply the needed power $gv_{c}$. It should therefore be no surprise that it is the slow time dependence of $V(q,t)$ in $H_{5}$ which makes the daemon stop driving when it runs out of $U$ `fuel'. And analogous phenomena will be observed in any analogous daemon system.

In fact the $H_{5}$ daemon must in general stop \textit{before} it consumes all its $U$. Whether because the washboard tilt $g$ increases, or because the washboard well depth $\kappa\sqrt{u(\Omega A-u)}$ eventually decreases, at some time $t_{0}$ there will no longer be any local minima in $V$. No bound orbits, and hence no steady downconversion, can continue past $t_{0}$, even if $U(t_{0})$ has not yet been drained to its ground state. At the minimum value $U=\Omega_{-}A$, the well depth is zero; but with finite $g$, $V$ will become a bumpy slope with no minima before the washboard completely disappears. A similar consideration will apply to any daemon whose effective Hamiltonian is adiabatically similar to $H_{5}$.
\subsection{B. Efficiency limits from adiabaticity}
Except for a zero-measure set of initial conditions, moreover, daemons must stop driving, and revert to adiabatic decoupling, even earlier than their $t_{0}$. This more stringent limit on Hamiltonian work extraction from high-frequency energy is the subtle inherent weakness in steady downconversion. The daemon can run for a long time, because it drains its fuel slowly: $\gamma\ll\omega$ is slow. But this means that, in the driving phase, the area $S$ that is enclosed in $(q,p)$ phase space by the periodic orbits of bound motion is an adiabatic invariant \cite{Goldstein}. At any $t<t_{0}$, any local well of $V$ will have a highest bound orbit, and the area $S_\mathrm{max}(t)$ which it encloses in the $(q,p)$ plane will be the largest $S$ that the local well can support at that time. For any steady downconversion trajectory, therefore, the orbit's enclosed area $S$ will remain constant (adiabatically, \textit{i.e.} up to small, fast oscillations), while the maximum area $S_\mathrm{max}$ which the well can support will eventually fall towards zero. After the time $t_{S}<t_{0}$, when $S_\mathrm{max}(t_{S})=S$ has fallen to $S$, the bound orbit with area $S$ can no longer continue, and the dynamical transition to adiabatic decoupling must occur. See Fig.~4.
\begin{figure}[t]\label{fig4}
\includegraphics[width=.45\textwidth]{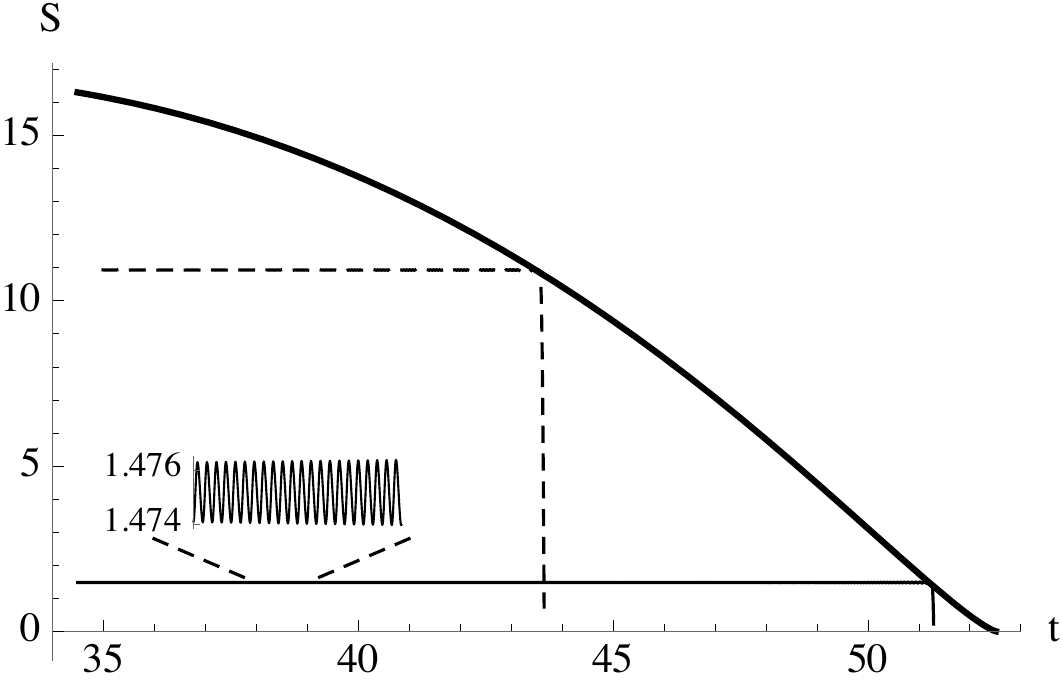}
\caption{Area $S(t)$ enclosed in $(q,p)$ phase space by bound orbits under $H_{5}$, for the parameter values of Fig.~1.\\ 
\textbf{Thick curve:} $S$ for the largest bound orbit that is possible at time $t$. This maximum area shrinks as the lattice tilt increases and the local minima become shallower. At time $t_{F}=52$ the harmonic force $-g(t)$ exceeds the maximum force the lattice can exert, and no minima exist. \\
\textbf{Solid curve:} $S$ for the trajectory shown in Fig.~1. Since the explicit time dependence of $H_{5}$ is slow, $S$ is adiabatically invariant; when the potential minimum becomes too shallow to contain it, the dynamical transition to adiabatic decoupling must occur. \\
\textbf{Dashed curve:} $S$ for a less efficient trajectory, under $H_{1}$ but with different initial conditions. Here the adiabatic invariance of $S$ forces the daemon to stop even though further driving is not forbidden by force limits or energy conservation. \\
\textbf{Inset:} The adiabatic invariant $S$ is not strictly constant, but exhibits small, rapid oscillations around an average value that holds steady over long times.}
\end{figure}

The adiabatic invariance of $S$ thus sets a limit on how much high-frequency energy a daemon can convert into work, which is in general lower (and as Fig.~4 shows, possibly much lower) than the limits set by energy conservation, or by the maximum force that can be exerted by the coupling term in $H_{1}$. This limitation is moreover inevitable, because it is inherent in the time scale hierarchy that defines a daemon, as sketched in Fig.~3. This generic daemon time scale hierarchy means that $S$ must be adiabatically invariant, and limit the duration of steady downconversion to be less than what would be allowed by conservation of energy alone.

\section{V. Outlook to Microthermodynamics}
The main result of this paper has been to prove by example that Hamiltonian daemons are in principle possible; small dynamical systems can sustain steady energy transfer from rapid motions to slow. We have also found a non-trivial limit on the amount of work that our example daemon can do, and as we have argued that our example daemon is a representative example of a large class of daemon systems, we have shown that this kind of efficiency limit is a generic feature of steady downconversion in small Hamiltonian systems. Small dynamical systems have traditionally been studied as toy models for more general phenomena, however. Since the behavior of daemons is at this level analogous to that of combustion engines, and since heat is after all fundamentally nothing but a certain rapid `kind of motion' \cite{Clausius}, we close our paper by discussing the possible implications of our results for the microscopic roots of thermodynamics.

\subsection{A. Daemons and thermodynamics}
The fact that Hamiltonian daemons can in general do less work than the limit allowed by energy conservation, because a certain phase space volume cannot decrease, is reminiscent of the thermodynamic limits on heat engine efficiency, which not only insist on conservation of total energy, but further say that not all energy is thermodynamically `available' for conversion into work, because entropy cannot decrease. In the statistical mechanics of the microcanonical ensemble, entropy is identified with (the logarithm of) a certain phase space volume (namely, that volume which is ergodically explored by the system as it undergoes chaotic evolution). It is therefore intriguing to speculate whether the mechanical constraints on Hamiltonian daemons, in which conservation of certain phase space volumes limits energy transfer, may perhaps be related somehow to the constraints of thermodynamics, whose microscopic origins remain obscure. 

Further investigations can moreover demonstrate additional analogies between daemon mechanics and thermodynamics, in the constraints on daemon ignition, and in entanglement growth in quantum mechanical daemons; these results are reported elsewhere \cite{paper2,paper3}. These analogies may even prompt an alternative hypothesis to the long-standing assumption that thermodynamics somehow `emerges', at mesoscopic scales, from non-thermodynamical microphysics. The alternative is that perhaps thermodynamics doesn't begin at mesoscopic scales, but simply persists into macroscopic scales from microscopic mechanics, where it already exists in embryonic form as the set of dynamical constraints upon steady downconversion.

On the other hand we must note that all the constraints on Hamiltonian daemons obtain within closed-system Hamiltonian mechanics, even without chaos or disorder. While the constraints on daemons may therefore resemble thermodynamics in their practical consequences, the reasons why the constraints apply would seem to be different. For the constraint that we have identified in this paper, for example, the difference between statistical mechanical entropy and the adiabatic invariant $S$ is drastic in a basic respect: the phase space area involved in limiting daemons does \textit{not} represent a volume that is ergodically explored by the system. The system's orbit under $H_{5}$ \textit{encloses} the area $S$, but it does not even enter the area $S$, let alone ergodically fill it. The concept of ergodic filling is fundamental in statistical mechanics, and if ergodic filling is really irrelevant to the mechanical constraints on steady downconversion, then the relationship between limits on daemons and thermodynamics may go beyond statistical mechanics itself, and lead either to an extension of thermodynamics into smaller systems (`microthermodynamics'), or the embedding of thermodynamics within a larger class of analogous constraints.

\subsection{B. Daemons and baths}
Should we really be certain, in any case, that statistical mechanics is the whole story of how thermodynamics emerges from mechanics? Statistical mechanics is based on introducing ensembles to provide `coarse-grained' descriptions, which contain no information about any aspects of time evolution that are rapidly ergodic. In this sense the first question that statistical mechanics addresses is about the exchange of \textit{information} between fast and slow variables, and its first answer is that equilibration occurs: information can be effectively lost from the slow sector, even though the total evolution is Liouvillian (or in quantum mechanics, unitary). Questions about energy exchange are then most often treated in a second step, in the sense that one studies energy transport under local (but not global) equilibrium, or postulates `baths' or `reservoirs' as sinks or sources of energy that always remain in equilibrium.

Active research continues into how information can thus effectively be lost within deterministic time evolution, and the physics of information is a major topic in general. By no means do we question the profound importance of information in physics; but we would argue also for the alternative approach, of considering energy exchange as the first question, instead of equilibration. Thermodynamics is after all a practical discipline, and engines do not exist to equilibrate, but to perform large amounts of steady work from compact fuel.

There is a basic physical reason why extracting large amounts of work from compact fuel demands downconversion: the units of energy are Mass $\times$ (Length)$^{2}$ / (Time)$^{2}$, and so a source of energy which is rich in comparison to its size and mass must possess a short time scale. It may therefore be worthwhile, as a kind of second front in the struggle to understand the microscopic roots of thermodynamics, to re-phrase the basic question. Instead of asking \textit{How can steady work be extracted from the equilibrated degrees of freedom of a heat bath?}, we can ask \textit{How can steady work be extracted from high-frequency degrees of freedom?} Further investigations of Hamiltonian daemons and related systems may thereby shed further light on thermodynamics.

Understanding Hamiltonian daemons may potentially even be practical. Chemical reactions are a ubiquitous source of this kind of dense energy, and the general problem of transferring their high-frequency energy into much slower motions is one that is not only solved macroscopically by engines, but also (somehow) by living organisms, on much smaller scales. Whether dynamical limitations on steady downconversion in small systems prove to be a microscopic limit of thermodynamics, or whether they turn out to be a parallel set of constraints, they may conceivably be of comparable importance to thermodynamics, for molecular machines \cite{biomotor2}.

Of course natural molecular machines are not isolated systems as daemons are, and any future artificial microscopic machines will also have to tolerate dissipation and noise if they are to be useful. We have only defined daemons as isolated, however, in order to ensure that they do not \textit{need} external power in order to operate. The adiabatic mechanisms by which daemons operate are normally quite robust, and so a generalized daemon, which was subject to dissipation and noise from its environment, might still very well function. If so, the processes of self-contained steady downconversion as by daemons, and work extraction from baths as by heat engines, may prove to be alternatives that are compatible --- or even to be alternative limits of a single phenomenon.

\section{Acknowledgements}
We thank our colleague Martin Strzys for suggesting `microthermodynamics' as a name for the hypothesis that thermodynamics extends, in some non-trivial form, to small systems. LG acknowledges funding from the German Excellence Initiative (DFG/GSC 266).

\begin{widetext}
\section{Appendix: Accuracy of $H_{5}$ as an adiabatic approximation to $H_{1}$}
\subsection{Non-resonant terms discarded to obtain $H_{2}$ from $H_{1}$}
A crucial step in our main text's chain of effective Hamiltonians $H_{1\to5}$ is discarding non-resonant terms from $H_{2}$. Without discarding them, the canonical transformation to Hamilton-Jacobi variables would instead have produced from $H_{1}$ the canonically equivalent Hamiltonian
\begin{eqnarray}\label{H15}
H_{1.5} &=& \frac{P^{2}}{2M}+\frac{M\nu^{2}}{2}Q^{2} + U - \kappa\sqrt{(\Omega_{+}A-U)(U-\Omega_{-}A)}\tilde{K}(\tau,\alpha,Q)\nonumber\\
\tilde{K}&\equiv& 4\cos(kQ)\cos(\Omega_{+}\tau+\alpha)\cos(\Omega_{-}\tau+\alpha)\equiv2\cos(kQ)\Bigl(\cos[(\Omega_{+}+\Omega_{-})\tau+2\alpha]+\cos(\Omega\tau)\Bigr)\nonumber\\
&\equiv&\cos[kQ+(\Omega_{+}+\Omega_{-})\tau+2\alpha]+\cos[kQ-(\Omega_{+}+\Omega_{-})\tau-2\alpha]+\cos(kQ+\Omega\tau)+\cos(kQ-\Omega\tau)\;.
\end{eqnarray}
The four terms in $\tilde{K}$ will then be resonant for $\dot{Q} \doteq \mp (\Omega_{+}+\Omega_{-})/k$ (first two terms) and $\dot{Q}\doteq \mp \Omega/k$ (last two terms), respectively. Non-resonant terms can be neglected, because of adiabatic decoupling; this is essentially the so-called ``rotating wave approximation'' of quantum optics. Since the only non-trivial motion that occurs in the trajectory shown in Fig.~1 involves $\dot{Q}\doteq v_{c}=\Omega/k$, the first three terms can be discarded, as far as this motion is concerned. This leaves us with $H_{2}$ from Eqn.~(3) of the main text. The accuracy of this rotating wave approximation, and of the other approximations which lead from $H_{1}$ to $H_{5}$, may then be assessed directly, by numerically solving the equations of motion for both Hamiltonians.

\subsection{Accuracy of $H_{5}$ as an approximation to $H_{1}$}
To compare evolution under the explicitly time-dependent adiabatic effective Hamiltonian $H_{5}$ to the full evolution under the original time-independent Hamiltonian $H_{1}$, we must relate the $q,p$ coordinates to the original coordinates. Note that $q$ and $p$ are exactly canonical, even though non-resonant terms have been dropped between $H_{1}$ and $H_{5}$ so that $H_{5}$ is only approximately equivalent to $H_{1}$. The momentum transformation is just a translation: $p=P-Mv_{c}=P-5000$. After inverting Eqn.~(2) to obtain $\tau$ in terms of $q_{\pm},p_{\pm}$ we find $q=Q-v_{c}\tau$ to be 
\begin{eqnarray}
q &=& - \frac{1}{k}\cos^{-1}\left(\frac{(\Omega_{+}\Omega_{-}q_{+}q_{-}+p_{+}p_{-})\cos(kQ)-(\Omega_{+}p_{+}q_{-}-\Omega_{-}p_{-}q_{+})\sin(kQ)}{\sqrt{(p_{+}^{2}+\Omega_{+}^{2}q_{+}^{2})(p_{-}^{2}+\Omega_{-}^{2}q_{-}^{2})}}\right)\;.
\end{eqnarray}\end{widetext}
where $\cos^{-1}$ is defined to be positive. (It turns out that in the driving phase $Q$ lags slightly behind $v_{c}\tau$, so that $q$ is negative; hence the overall minus sign.) From the above coordinate mappings we can therefore compute the exact $q(t),p(t)$ from the numerical solution to the full equations of motion for all six original canonical coordinates under $H_{1}$, as plotted in Fig.~1, and compare this to adiabatically approximate evolution of $q$ and $p$ alone under $H_{5}$. We will refer to these solutions as $(q_{1},p_{1})$ and $(q_{5},p_{5})$, respectively. 

To obtain $(q_{5},p_{5})$, we must first determine $H_{5}$ explicitly by supplying the correct slowly-varying external force. The initial conditions of Fig.~1 provide
\begin{eqnarray}
U(0)=\frac{m\Omega_{+}^{2}}{2}q_{+}^{2}(0)=1.2\times10^{8}\;.
\end{eqnarray}
The daemon begins steady downconversion at around $t=13$ --- the transition is by definition not something that can occur at one precise instant, because it is a change from non-periodic to periodic motion in the $(q,p)$ plane, and the exact moment within the first period at which the motion first becomes periodic is not well defined. We can, however, extrapolate the nearly constant slope of $Q(t)$, during the driving phase, back to the time $t=12.51$ at which $Q(t)$ would have been zero if the driving phase had begun from zero $Q$. We therefore deduce the adiabatically approximate
\begin{eqnarray}
U(t) \doteq \tilde{U}(t)\equiv U(0)-\frac{M\nu^{2}}{2}[v_{c}(t-12.51)]^{2}\;.
\end{eqnarray}
This yields
\begin{eqnarray}
H_{5}(q,p,t) &=& \frac{p^{2}}{2M}+ M\nu^{2}v_{c}(t-12.51)q\nonumber\\
&&-\kappa \sqrt{[\tilde{U}(t)-\Omega_{-}A][\Omega_{+}A-\tilde{U}(t)]}\cos(kq)\;.\nonumber
\end{eqnarray}

Adiabatic effective Hamiltonians like $H_{5}$ are in general determined by decomposing the exact evolution into a sum of slow and fast parts: $q(t)=q_{s}(t)+q_{f}(t)$, for example. Where adiabatic methods are warranted, this decomposition remains valid over long times, and the fast term $q_{f}$ remains small in amplitude, as well as high in frequency, whereas the slow part $q_{s}$ may slowly change by large amounts. The goal of adiabatic methods is to discard the fast parts like $q_{f}$ and describe $q_{s}$ accurately, over long times, with simpler equations; even with ample experience it is astonishing how well this can work. A subtlety which is sometimes overlooked in deriving adiabatic methods, however, is that the initial conditions which must be used for the adiabatic approximate evolution, in order to match the slow part of the exact motion, are generally not the same as the exact initial conditions, because the exact initial conditions represent the sum of the initial values of both fast and slow components. Slightly perturbed or renormalized initial conditions must therefore be used in the adiabatic approximation. Since the full and exact initial conditions determine the entire evolution, the correct adiabatic initial conditions can be derived as particular functions of the exact initial conditions; but it is usually easier simply to determine the adiabatic initial conditions by fitting the first few periods of adiabatic evolution to the exact evolution. The magic of adiabaticity is that this initially fitted approximation will then remain accurate for a long time.

For the evolution shown in Fig.~1 we solve the canonical equations of motion derived from $H_{5}$, from the initial time $t=16$, shortly after the daemon's driving phase has begun. (The adiabatic method does not work as well right around the dynamical transition, where post-adiabatic effects are significant and other methods must be applied.) By tuning the initial values to $q_{5}(16)=-0.01$ and $p_{5}(16)=-100$, we can achieve the good agreement shown in Fig.~A1 between $H_{5}$ evolution and the exact $H_{1}$ evolution with the initial conditions of Fig.~1 (which applied at $t=0$), over the early time window $16\leq t\leq 16.4$.
\begin{figure}\label{A1}
\includegraphics[width=.45\textwidth]{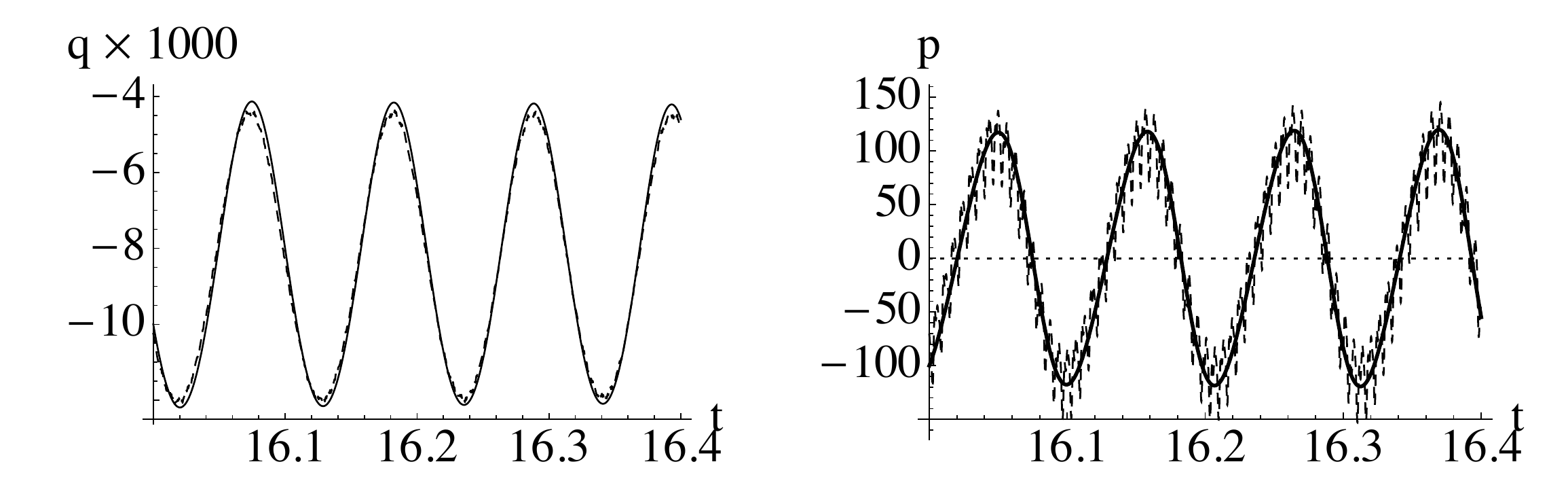}
\caption{Adiabatic evolution $(q_{5}(t),p_{5}(t))$ (solid curves) versus exact evolution $(q_{1}(t),p_{1}(t))$ (dashed curves), over the interval $16\leq t\leq 16.4$. Initial conditions for the exact evolution are those stated in the caption of Fig.~1 in the main text, and apply at $t=0$. Initial conditions for the adiabatic evolution are $q_{5}(16)=-0.01, p_{5}(16)=-100$. Fitting was done simply by eye. The suppression of high-frequency components from the adiabatic evolution is not error: it is what the adiabatic evolution is supposed to do.}
\end{figure}

We can then see how well this adiabatic solution $q_{5},p_{5}$ tracks the exact evolution $q_{1},p_{1}$ at later times. To do this, we are not allowed to retune the initial conditions for each later time period; we must keep the same initial conditions at $t=16$, and see what they imply for $q_{5},p_{5}$ later. Adiabatic evolution is not normally good, however, at following the exact phase of periodic motion over long times. Time translation is a zero mode, and the time coordinates of the exact and approximate solutions tend to drift slowly out of synchronization. We can therefore compare $(q_{1}(t),p_{1}(t))$ with $q_{5}(t-\Delta t),p_{5}(t-\Delta t))$, with $\Delta t$ a new fitting parameter than is allowed to vary slowly in time. The agreement in phase that we then obtain, between exact and adiabatic solutions, is merely due to fitting; but the agreement in amplitude, and in all other slow features of the periodic motion, is non-trivial --- indeed, remarkable. Over the long times that we investigate, the amplitude and frequency of both slow and fast components of the motion all change substantially, but the adiabatic approximation tracks the slow amplitudes and frequencies extremely well. See Fig.~A2.

\begin{figure}\label{A2}
\includegraphics[width=.45\textwidth]{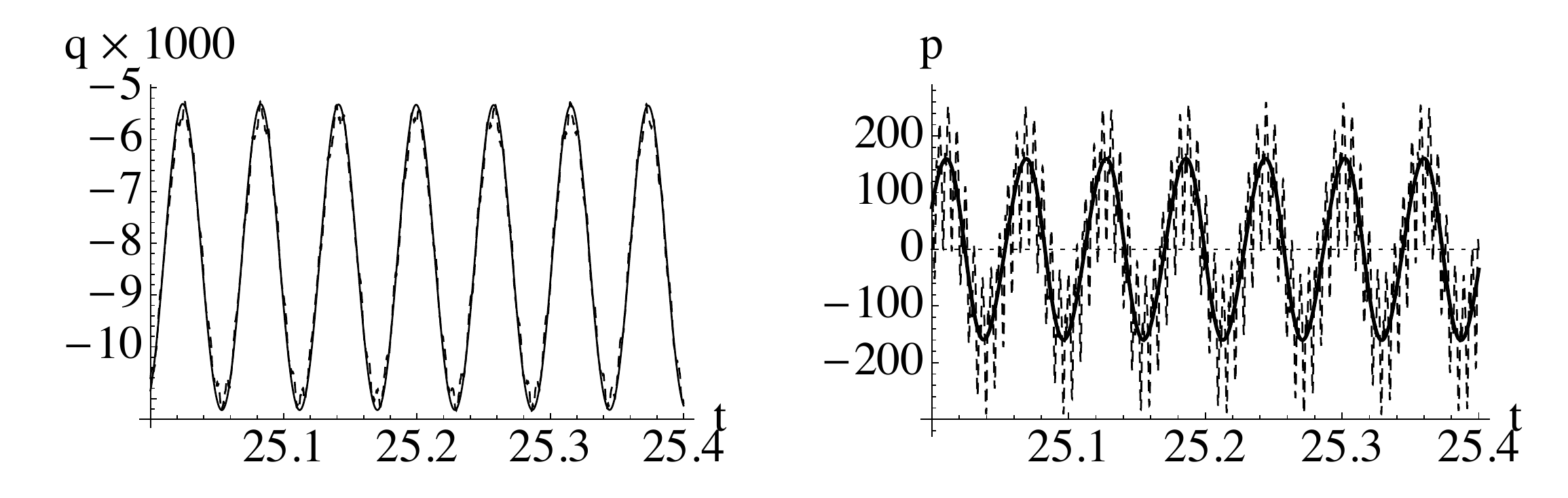}
\includegraphics[width=.45\textwidth]{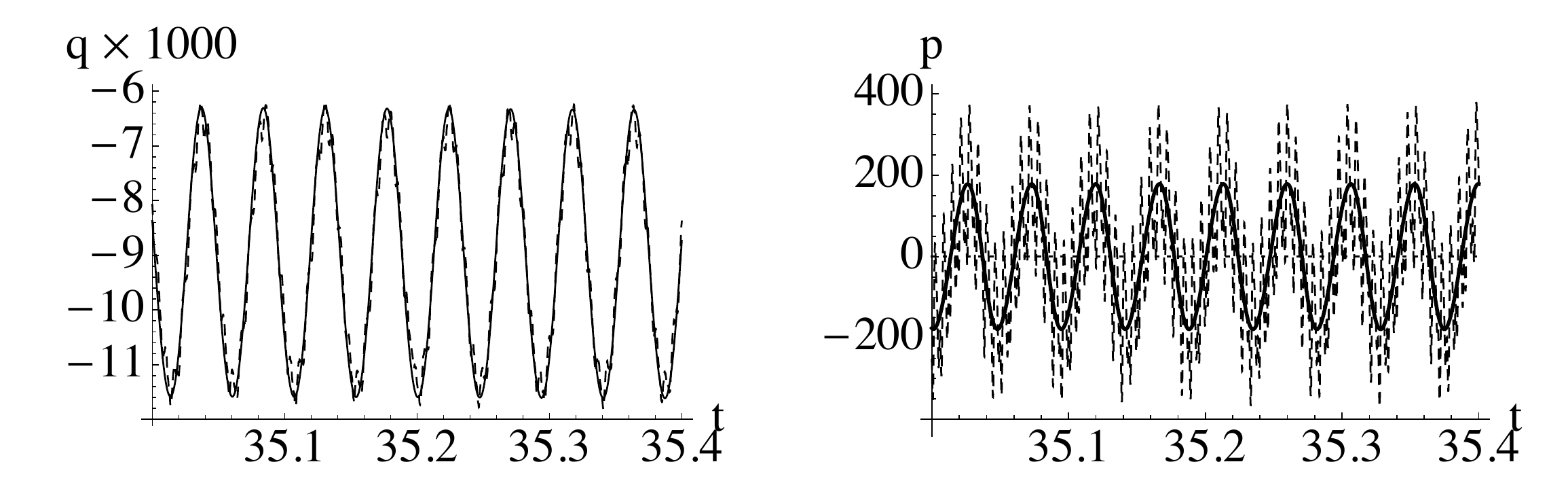}
\includegraphics[width=.45\textwidth]{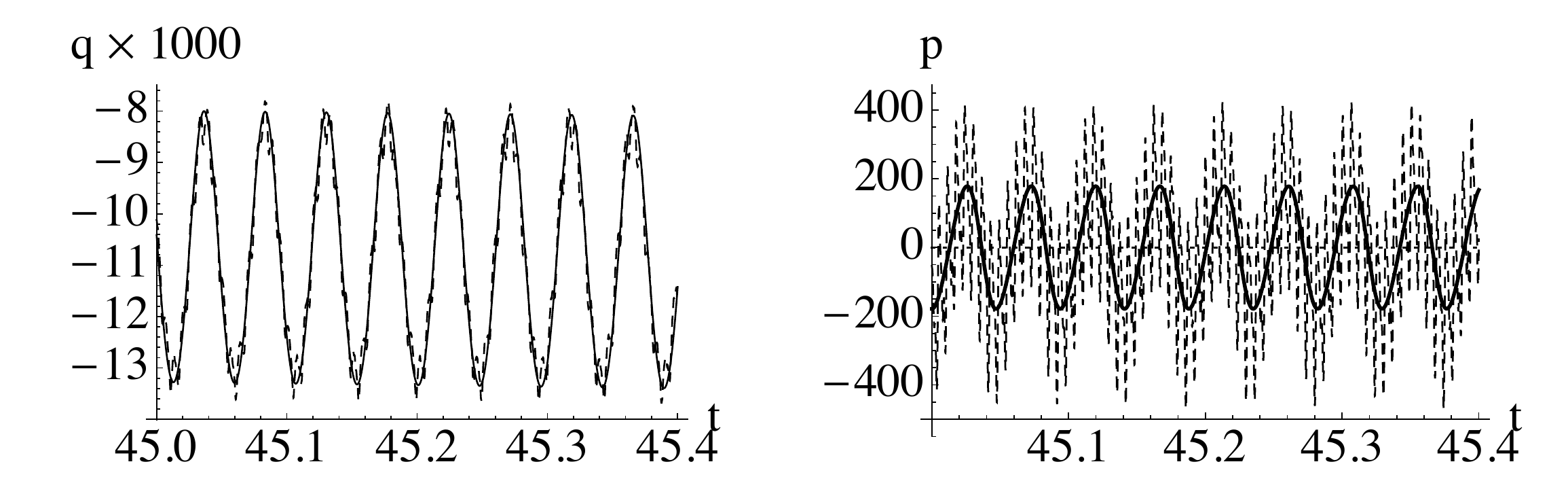}
\caption{Same as Fig.~A1 above, but for time intervals $[25,25.4], [35,35.4]$ and $[45,45.4]$, respectively, from top to bottom. Solid curves are the same adiabatic solution plotted in Fig.~A1, but with fitted time shifts $\Delta t = -0.001$, $+0.025$, and $-0.001$, from top to bottom.}
\end{figure}

\subsection{Adiabatic invariance of $S$ under $H_{5}$ evolution}
As well as being derived from $H_{1}$ as an adiabatic effective Hamiltonian, $H_{5}$ is itself a good target for adiabatic theory, because its explicit dependence on $t$ is very slow compared to the frequency of the bound orbits it describes during steady downconversion. The slowness of the $t$-dependence of the functional form of $H_{5}$ naturally suggests an approximation based on the fact that $H_{5}$ remains \emph{nearly} the same function over several periods. And indeed one can confirm that the instantaneous value $E_{5}(t)=H_{5}(t)$ for a particular trajectory changes only slightly, over several periods. The changes in $E_{5}(t)$ do not remain small over long times, however; $E_{5}(t)$ changes slowly, but it slowly changes by a large amount. When a Hamiltonian like $H_{5}$ is a slowly time-dependent function, the adiabatic theorem states that the quantity which remains nearly constant over long times is not the value of the Hamiltonian, but rather the \emph{action} $S(E_{5},t)$. 

The action $S$ is defined for any fixed $t$ and $H_{5}$ value $E_{5}$ to be the area in the $(q,p)$ phase space plane which is enclosed by the closed curve $q_{5}(s),p_{5}(s)$ which satisfies $H_{5}(q_{5},p_{5},t)=E_{5}$. In our main text this action is computed and plotted in Fig.~4, for two different exact trajectories under $H_{1}$. For each trajectory the corresponding adiabatic $q_{5}(t),p_{5}(t)$ are determined by initial fitting, as described above. For each of the two trajectories shown in Fig.~4, the value $E_{5}(t)=H_{5}(t)$ is computed from the initially fitted $q_{5}(t),p_{5}(t)$, and then the corresponding enclosed phase space area $S$ is computed by numerical integration. (No subsequently fitted time shifts $\Delta t$ are employed in these computations, because $E_{5}(t)$ is a slowly varying function that is not sensitive to short time translations.) The plotted $S(t)$ are indeed close to being constant, until the transition to adiabatic decoupling occurs as explained in the main text. 

As the inset in Fig.~4 shows, however, $S$ is not an exact constant of the motion. As an adiabatic invariant, it exhibits small and fast oscillations around an average value that remains constant over a long time. For this reason the time at which steady downconversion ceases is not only intrinsically ill-defined within a period; it is impossible to predict from slow variables alone, because once the bound motion is only barely bound, a small post-adiabatic correction may be enough to let the system escape across the adiabatic $H_{5}$ separatrix into unbound motion. This adiabatic uncertainty in the precise time at which the daemon will stop working can easily be a very small fraction of the total duration of steady downconversion, however. For daemons whose time scale ratios $\Omega/\omega$ and $\omega/\gamma$ are large, the significance of post-adiabatic corrections compared to adiabatic results may be comparable to the importance of thermal fluctuations relative to equilibrium thermodynamics in macroscopic engines.

\begin{figure}\label{A3}
\includegraphics[width=.45\textwidth]{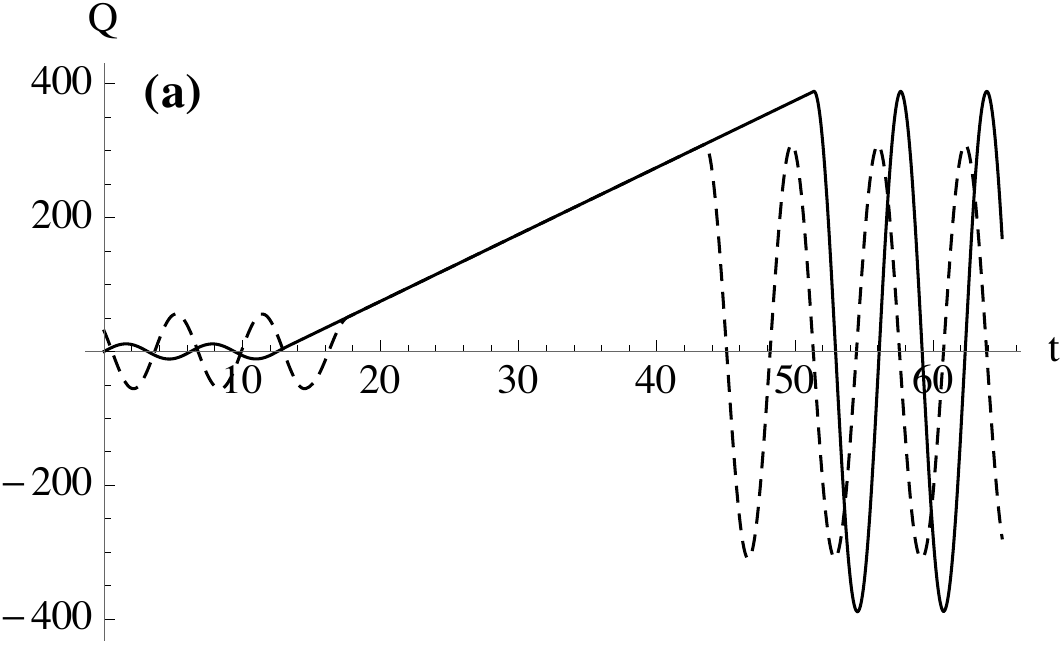}
\includegraphics[width=.4\textwidth]{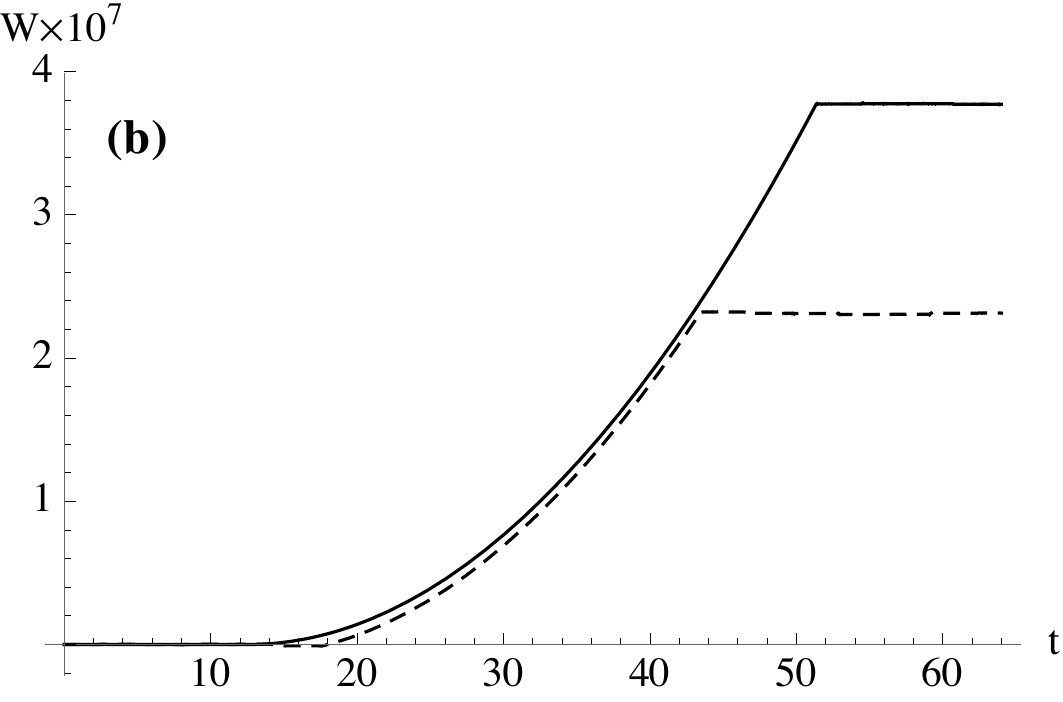}
\caption{\textbf{(a)} Same as Fig.~1a) of main text, but with the dashed trajectory from main text Fig.~4 for comparison. When both trajectories are in the driving phase, they are impossible to distinguish on this scale; this is a consequence of how the second trajectory was selected, as explained in the Appendix text. The dashed trajectory ceases steady downconversion at $t\doteq 43.7$, in agreement with Fig.~4.\\
\textbf{(b)} Work $W(t)=E_{S}(t)-E_{S}(0)$ done by the daemon on the slow mass, where $E_{S}=P^{2}/(2M)+(M\nu^{2}/2)Q^{2}$ is the energy of the slow oscillator. Final values $W(65)$ are $3.77\times 10^{7}$ (solid) and $2.31\times10^{7}$ (dashed).}
\end{figure}

The dashed line in the main text's Fig.~4 represents the same function $S$, computed for a slightly different trajectory $q'_{5}(t),p'_{5}(t)$, corresponding with the same adiabatic kind of accuracy to a different exact trajectory $q'_{1}(t),p'_{1}(t)$. To determine the particular different trajectory, a certain fine-tuning of initial conditions was employed for graphical reasons. The thick curve $S_{\mathrm{max}}(t)$ depends on $J$, and hence even though it is defined by the adiabatic Hamiltonian $H_{5}$, it really depends on the full initial conditions of the six canonical co-ordinates in the exact $H_{1}$. We were able to generate a second exact trajectory with (almost) exactly the same $S_{\mathrm{max}}(t)$ as the trajectory from Fig.~1, so that it could be plotted in the same Fig.~4, by imposing initial conditions at time $t_{I}'=35.0002$. The conditions we imposed for the new trajectory at this time were that all six phase-space variables should equal those of the Fig.~1 trajectory at the same time $t_{I}'$, except that we set $P'(t_{I}')=P(t_{I}')-300$. This displacement of $P$, while keeping all else fixed, produced bound motion, as in the Fig.~1 trajectory, but with higher amplitude oscillation in the local well of $V(q,t)$; hence higher $S$, and so earlier stalling time $t_{S}\doteq 43.7$.

The dashed trajectory of Fig.~4 has only slightly higher $E_{5}$ than the solid trajectory; but this small difference eventually makes a large difference. During the driving phase of motion, the dashed trajectory of $Q(t)$ as shown in Fig.~A3a) does not visibly differ from the original one, because the difference is only in oscillations to small to be seen on the plotted scale. The slightly higher $E_{5}$ in the dashed case makes a substantial difference in the amount of work the daemon does, however, because the higher $E_{5}$ of the dashed trajectory means higher $S$ and earlier $t_{S}$, so that steady downconversion stops sooner in the dashed trajectory, as we see in both parts of Fig.~A3, as well as in the main text's Fig.~4. In the dashed case the daemon does nearly 39\% less total work on the slow mass than in the original trajectory of Fig.~1. The general limitation on daemon efficiency from the adiabatic invariance of $S$ is not crippling, but neither is it trivial.

\begin{thebibliography}{10}
\bibitem{explains}This explains the slight anharmonic perturbations in $\dot{Q}(t)$ which can be seen in the first 10 time units in Fig.~1, whenever $\dot{Q}\sim\pm 10$.
\end{thebibliography}
\end{document}